\documentclass[%
twocolumn,
showpacs,preprintnumbers,
nofootinbib,
nobibnotes,
bibnotes,
amsmath,amssymb,
aps,
prd,
]{revtex4}

\usepackage{graphicx}
\usepackage{dcolumn}
\usepackage{bm}
\usepackage{float}

\begin{document}

\title{Warm inflation with a generalized Langevin equation scenario}

\author{Xi-Bin Li}
\email{lxbbnu@mail.bnu.edu.cn}
\affiliation{College of Physics and Electronic Information, Inner Mongolia Normal University, Huhhot 010022, China}
\affiliation{School of Physics and Technology, Wuhan University, Wuhan 430072, China}
\affiliation{Department of Physics, Beijing Normal University, Beijing 100875, China}

\author{Shi-Wei Yan}
\email{yansw@bnu.edu.cn}
\affiliation{Department of Physics, Beijing Normal University, Beijing 100875, China}

\author{He Wang}
\email{hewang@mail.bnu.edu.cn}
\affiliation{Department of Physics, Beijing Normal University, Beijing 100875, China}

\author{Jian-Yang Zhu}
\thanks{Corresponding author}
\email{zhujy@bnu.edu.cn}
\affiliation{Department of Physics, Beijing Normal University, Beijing 100875, China}

\date{\today}

\begin{abstract}
  In this paper, we discuss the warm inflation model with both a Langevin equation and a generalized Langevin equation scenario. As a brief picture to illustrate the basic properties of stochastic differential equation in warm inflation, this paper is started from a simple condition with constant dissipative coefficient. In this model, we prove the perturbed inflaton field exhibits a stationary process on large scale, so the perturbed field has a scale-invariant power spectrum. Then we study the warm inflation with a generalized Langevin equation scenario. The perturbed field in such model also shows a stationary process and the power spectrum is quite similar to the one in cold inflation. If choosing an appropriate fluctuation-dissipation relation, we can get a spectrum same as the cold inflation. In a word, we attempt to show the rationality of warm inflationary scenario via statistical physics method.
\end{abstract}

\maketitle


\section{Introduction}
\label{introduction}

Warm inflation model was established as a candidate scenario to overcome some defects in cold inflation \cite{doi:10.1142/S0217751X09044206,BARTRUM2014116}.
However, it was realized a few years after its original proposal that the idea of warm inflation was not easy to be realized in concrete models and even is simply not possible in relevant works \cite{PhysRevD.60.083509,PhysRevLett.83.264}. Some problems mentioned were suspected in such scenario. Shortly afterwards many successful models of the warm inflation have been established, in which the inflaton indirectly interacts with the light degrees of freedom though a heavy mediator fields instead of being coupled with a light field directly \cite{doi:10.1142/S0217751X09044206,PhysRevD.84.103503,PhysRevLett.117.151301,1475-7516-2011-09-033}. The evolution of the inflaton field can be properly determined in the context of the in-in, or the Schwinger closed-time path functional formalism \cite{rammer2007quantum}. This equation not only displays both dissipation and non-Markovian stochastic noise terms, but also can be regarded as a generalized Langevin-like equation of motion \cite{PhysRevD.76.083520,PhysRevD.91.083540}.

Compared with the predictions of the cold inflation that primordial density fluctuations mostly from quantum fluctuation and thermal bath are only generated at the end of inflation \cite{RevModPhys.78.537}, the warm inflation model suggests that our universe is hot during the whole inflation when inflaton fields couple with the thermal bath and the primary source of density fluctuations comes from thermal fluctuations \cite{PhysRevD.64.063513, PhysRevD.62.083517, PhysRevD.69.083525}. The equation of motion for warm inflation can be written as a stochastic Lengevin equation, in which there is a dissipation term to describe the inflaton fields coupling with the thermal bath and there is also a fluctuation term described by a stochastic noise term \cite{PhysRevD.71.023513, PhysRevD.76.083520}. The fundamental principles of the warm inflation have been described and reviewed in \cite{PhysRevD.50.2441}.

The Langevin equation is widely used in the dynamics of a Brownian particle in phase space which is described by the Markovian set of the differential equations \cite{keizer2012statistical,reichl2016modern,p2012langevin}. If we consider the dissipative coefficient as a constant, this stochastic differential equation can be casted as the Langevin equation which represents a Markov process. Otherwise if the dissipative coefficient is a function (called integral kernel, damping kernel or memory kernel), the stochastic equation, as a differential integral equation, can be called a generalized Langevin equation corresponding to a non-Markov process \cite{OLIVARESRIVAS201676,PhysRevA.26.1589}. Both the dissipative coefficient and integral kernel, obviously, yield the fluctuation-dissipation theory \cite{klages2008anomalous}.

In this paper, we attempt to illustrate the rationality of explanation with the warm inflationary scenario via statistical physics method.
To achieve that goal, we need to prove the process described by the (generalized) Langevin equations are stationary process on large scale.
Thus we can examine the scale-invariant power spectrum for the reason that a stationary process means invariant expectation variance \cite{prabhu2007stochastic}, by which way we can get the power spectrum at the horizon-crossing scale. The spectrum from the Langevin scenario is the same as the one via the Green's function \cite{PhysRevD.62.083517}, while the spectrum from the generalized Langevin scenario is similar to the cold inflation. The isotropy and homogeneity of cosmic microwave background shows a near thermal equilibrium state must be hold for our early universe \cite{2016A&A...594A..13P, WMAP}. Based on this observational fact, we get the freeze-out wave number and the approximate conditions satisfied in warm inflationary model.

This paper is organized as follows: In Sec.~\ref{Langevin_equation_WI}, we first give a brief introduction to Langevin equations and warm inflation together with their properties, then we study the warm inflation with a Langevin equation scenario and get some suggestive results to prepare further discussions in next section. In Sec. \ref{generalized_Langevin_equation_WI}, we study the warm inflation with generalized Langevin equation scenario and get some important conclusions. Finally, in Sec. \ref{conclusion}, we conclude our work and give some further discussions about our results.

\section{\label{Langevin_equation_WI}Warm inflation with constant dissipative coefficient}

\subsection{Langevin equation and warm inflation}
\label{LE_and_WI}

Before starting the discussion to the thermodynamic properties of the warm inflation, it's necessary to have a brief introduction to several thermodynamics foundations. In physics, the Langevin equation is a stochastic differential equation describing the statistical properties of particles with irregular motion. The Langevin dynamics method has the form \cite{p2012langevin}
\begin{eqnarray}
    m\dot{v}+\beta v+U'(x)=\xi(t), \label{LE}
\end{eqnarray}
where $\beta$ is a constant (called dissipative coefficient) which describes the damped effect of a particle coupling with other particles and $\xi(t)$ is a stochastic force which denotes the fluctuate effect of a particle driven by other particles nearby.
The dissipative constant and fluctuate force follow the fluctuation-dissipation theorem
\begin{eqnarray}
    \langle\xi(t)\xi(t')\rangle=mk_\text{B}T\beta\delta(t-t'). \label{FDT}
\end{eqnarray}
The stochastic differential equation of Eq.~(\ref{LE}) describes a Markov progress, which means the stochastic properties of a thermodynamics system at time $t$ are independent on its previous time $t'<t$.
If a stochastic variable depends on its previous state, this variable is called a non-Markov progress which has the form
\begin{eqnarray}
    m\dot{v}(t)+m\int^t_0{\gamma(t-t')v(t')\text{d}t'}+U'(x)=\xi(t), \label{GLE}
\end{eqnarray}
where $\gamma(t-t')$ is called damped kernel and $\xi(t)$ is also named stochastic force.
The damped kernel and stochastic force, obviously, follow the fluctuation-dissipation theorem \cite{klages2008anomalous}:
\begin{eqnarray}
    \langle\xi(t)\xi(t')\rangle=mk_\text{B}T\gamma(|t-t'|). \label{GFDT}
\end{eqnarray}
The stochastic differential integrate equation~(\ref{GLE}) is a non-Markov progress.

In warm inflation model, the equation of motion of background field is often written as the Langevin equation
 \begin{eqnarray}
    \Big[\frac{\partial^2 }{\partial t^2}+(3H+\Gamma)\frac{\partial}{\partial t}-\frac{1}{a^2}\nabla^2\Big]\Phi+\frac{\partial V(\Phi)}{\partial \Phi} = \xi(\textbf{x},t), \label{EOMwarminflaton}
\end{eqnarray}
where $\Gamma$ is the dissipation coefficient and $\xi$ is the thermal noise fluctuation.
In this paper, we consider only in the case of de Sitter space-time, where $a(t)=\textrm{exp}(Ht)$ and a constant $H$.
According to the fluctuation-dissipation theorem, dissipation coefficient $\Gamma$ and fluctuation noise $\xi$ have the relation
\begin{eqnarray}
    \langle \xi(\textbf{x},t)\xi^*(\textbf{x},t')\rangle=2\Gamma T a^{-3}\delta^3(\textbf{x}-\textbf{x}')\delta(t-t'). \label{dis_flu_relation}
\end{eqnarray}
The Fourier transformation of Eq.~(\ref{dis_flu_relation}) is
\begin{eqnarray}
    \langle \xi(\textbf{k},t)\xi^*(\textbf{k}',t')\rangle=2(2\pi^3)\Gamma T a^{-3}\delta^3(\textbf{k}-\textbf{k}')\delta(t-t'). \nonumber\\ \label{dis_flu_relation_k}
\end{eqnarray}
Usually $\Gamma$ is a function of both background homogeneous inflaton field $\Phi$ and temperature $T$ \cite{1475-7516-2011-09-033}.

The inflaton field operator $\Phi(\textbf{x},t)$ is often separated into the parts as follow
\begin{eqnarray}
   \Phi(\textbf{x},t)=\phi(t)+\delta\varphi(\textbf{x},t), \label{inflaton}
\end{eqnarray}
where $\delta\varphi(\textbf{x},t)$ is the perturbed part of inflaton, and $\phi(t)$ is the background homogeneous inflaton field defended as
\begin{eqnarray}
   \phi(t)=\frac{1}{\Omega}\int_\Omega{\text{d}^3x \Phi(\textbf{x},t)}. \label{BGinflaton}
\end{eqnarray}
Here, $\Omega$ is particle horizon size $\Omega=1/H$.
With this relation, Eq.(\ref{EOMwarminflaton}) reads
\begin{eqnarray}
     \frac{\partial^2 \phi}{\partial t^2}+[3H+\Gamma]\frac{\partial \phi}{\partial t}+V,_\phi(\phi) &=& 0,\label{EOM_BGinflaton} \\
    \Big\{\frac{\partial^2}{\partial t^2}+[3H+\Gamma(\phi)]\frac{\partial}{\partial t}+\frac{k^2}{a^2}+\nonumber \\
    \Gamma_\phi(\phi)\dot{\phi}
    +V_{\phi\phi}(\phi)\Big\}\delta\varphi &=& \xi(\textbf{k},t). \label{EOM_pertubed_field}
\end{eqnarray}

It is also necessary to define some slow-roll parameters for warm inflation,
\begin{eqnarray}
    \varepsilon &=& \frac{1}{16 \pi G}\Big(\frac{V,_\phi}{V}\Big)^2 \ll 1+r , \label{varepsilon} \\
    \eta &=& \frac{1}{8 \pi G}\frac{V,_{\phi\phi}}{V} \ll 1+r , \label{eta}
\end{eqnarray}
and
\begin{eqnarray}
    \beta &=& \frac{1}{8 \pi G}\frac{ \Gamma,_\phi V,_{\phi}}{\Gamma V} \ll 1+r , \label{beta}
\end{eqnarray}
where $r$ is the ratio between the dissipation coefficient $\Gamma$ and the Hubble parameter $H$, i.e., $r\equiv\Gamma/3H$.

\subsection{Power Spectrum}

We first consider the condition of dissipative coefficient with a constant.
Although the dissipative term may be very complex as a function of inflationary fields or cosmic time $t$,
such a simple model would help us to get several preliminary conclusions and illustrate some useful properties of warm inflationary scenario.

With the slow-roll approximation, we treat the dissipative coefficient as a constant so that the Langevin equation~(\ref{EOM_pertubed_field}) approximately reads \cite{2000NuPhB.585..666B}
\begin{eqnarray}
    (3H+\Gamma)\frac{\text{d}\delta\varphi(\textbf{k},t)}{\text{d}t}+[k^2_p+V''(\phi)]\delta\varphi(\textbf{k},t) \approx \xi(\textbf{k},t), \nonumber \\ \label{approx_EOM}
\end{eqnarray}
where $V''(\phi)=\text{d}^2V(\phi)/\text{d}\phi^2$ and $\phi$ is defined in Eq.(\ref{BGinflaton}), $k_p=k/a$ is the physical wave number and $k$ is the conformal wave number. The process in the equation above is a stationary, Markov, and Gaussian process. The solution of Eq.~(\ref{approx_EOM}) is
\begin{eqnarray}
    \delta\varphi(\textbf{k},t)\approx & &\frac{1}{3H+\Gamma}e^{-(t-t_0)/\tau(\phi)}\delta\varphi(\textbf{k},t_0) \nonumber\\
    & &+\frac{1}{3H+\Gamma}\int^t_{t_0}{\text{e}^{-(t-t_0)/\tau(\phi)}\xi(\textbf{k},t')\text{d}t'}. \label{approx_solution}
\end{eqnarray}
where $t_0$ is any coordinate time during inflation and
\begin{eqnarray}
\tau (\phi )=\frac{3H+\Gamma }{k_p^2+V^{\prime \prime }(\phi )}. \label{relaxation time}
\end{eqnarray}
In statistical mechanics, $\tau(\phi)$ is called relaxation time which means a time during which a thermodynamic system returns from a perturbed state into the equilibrium state. The observation of isotropy and homogeneous cosmic micro background, of cause, requires that the relaxation time must be much smaller than the cosmic time, i.e., $\tau(\phi)\ll 1/3H$, which yields
\begin{eqnarray}
    k_p\gg 3H(1+r)^{1/2}, \label{approx_of_equilibrium}
\end{eqnarray}
where we have used the relation of slow-roll parameter $V''/H^2=3\eta\ll 1+r$.
Define the freeze-out number
\begin{eqnarray}
    k_F\equiv H(1+r)^{1/2}. \label{freeze_out_number}
\end{eqnarray}
From Eq.~(\ref{freeze_out_number}), it's easy to find the freeze-out number in warm inflation degenerates to that in cold inflation with weak dissipative condition while it approximately equals to $(H\Gamma)^{1/2}$ with strong dissipative condition $\Gamma\gg 3H$.
In warm inflation, the freeze-out wave number is always larger than the Hubble crossing wave number $k=aH$, which means the freeze-out time will always precede the Hubble crossing time. The evolution of the inflaton is mainly deterministic during time $t>t_F$.

The autocorrelation function of the perturbed inflation field is
\begin{eqnarray}
&&\{\langle \delta \varphi ({\bf k},t_1)\delta \varphi ^{*}({\bf k}^{\prime
},t_2)\rangle \}  \nonumber \\
&=&\delta \varphi ({\bf k},t_0)\delta \varphi ({\bf k}^{\prime },t_0)\text{e}%
^{-(t_1+t_2)/\tau (\phi )}  \nonumber \\
&&+\frac{2(2\pi )^3k_{\text{B}}T\Gamma \delta ^3({\bf k}-{\bf k}^{\prime })}{%
(3H+\Gamma )^2}\text{e}^{-(t_1+t_2)/\tau (\phi )}  \nonumber \\
&&\quad \quad \times \int_{t_0}^{t_1}\int_{t_0}^{t_2}{\text{e}%
^{(s_1+s_2)/\tau (\phi )}a^{-3}(s_1)\delta (s_1-s_2)\text{d}s_1\text{d}s_2} \nonumber \\
\label{correlation1}
\end{eqnarray}
where Eq.~(\ref{dis_flu_relation_k}) has been used, and $\langle \cdot \cdot \cdot
\rangle $ denotes stochastic averaging while $\{\cdot \cdot \cdot \}$ denotes
the stochastic averaging on initial state $\delta \varphi ({\bf k},t_0)$.
The double integral in Eq.(\ref{correlation1}) contains a $\delta $ function,
so we need to integrate first to the lager one in $t_1$ and $t_2$. Thus,
\begin{widetext}
\begin{align}
&\left\langle \delta \varphi ({\bf k},t_1)\delta \varphi ^{*}({\bf k}%
^{\prime },t_2)\right\rangle   \nonumber \\
&=\left\{ \delta \varphi ({\bf k},t_0)\delta \varphi ({\bf k}^{\prime
},t_0)\right\} \text{e}^{-(t_1+t_2)/\tau (\phi )}+\frac{2(2\pi )^3k_{\text{B}%
}T\Gamma \delta ^3\left( {\bf k}-{\bf k}^{\prime }\right) }{(3H+\Gamma )^2}
\nonumber \\
&\qquad \times \text{e}^{-(t_1+t_2)/\tau (\phi )}\int_{t_0}^{\text{min}%
(t_1,t_2)}\int_{t_0}^{\text{max}(t_1,t_2)}{\text{e}^{(s_1+s_2)/\tau (\phi
)}a^{-3}(s_1)\delta (s_1-s_2)\text{d}s_1\text{d}s_2}  \nonumber \\
&=\left\{ \delta \varphi ({\bf k},t_0)\delta \varphi ^{*}({\bf k}^{\prime
},t_0)\right\} \text{e}^{-(t_1+t_2)/\tau (\phi )}+\frac{2(2\pi )^3k_{\text{B}%
}T\Gamma \delta ^3\left( {\bf k}-{\bf k}^{\prime }\right) }{(3H+\Gamma )^2}%
\text{e}^{-(t_1+t_2)/\tau (\phi )}\int_{t_0}^{\text{min}(t_1,t_2)}{\text{e}^{%
{2s}/{\tau (\phi )}}\text{e}^{-3Hs}\text{d}s}  \nonumber \\
&=\left\{ \delta \varphi ({\bf k},t_0)\delta \varphi ^{*}({\bf k}^{\prime
},t_0)\right\} \text{e}^{-(t_1+t_2)/\tau (\phi )}+\frac{(2\pi )^3k_{\text{B}%
}T\Gamma \tau (\phi )\delta ^3\left( {\bf k}-{\bf k}^{\prime }\right) }{%
(3H+\Gamma )^2\left( 2-3H\tau (\phi )\right) }  \nonumber \\
&\qquad \times \left[ \text{e}^{-[t_1+t_2-2\text{min}(t_1,t_2)]/\tau (\phi
)}a^{-3}(t_m)-\text{e}^{-(t_1+t_2)/\tau (\phi )}a^{-3}(t_0)\right]
\nonumber \\
&=\left\{ \delta \varphi ({\bf k},t_0)\delta \varphi ^{*}({\bf k}^{\prime
},t_0)\right\} e^{-(t_1+t_2)/\tau (\phi )}+\frac{(2\pi )^3k_{\text{B}%
}T\Gamma \tau (\phi )\delta ^3\left( {\bf k}-{\bf k}^{\prime }\right) }{%
(3H+\Gamma )^2\left( 2-3H\tau (\phi )\right) }a^{-3}\left[ \text{e}%
^{-|t_1-t_2|/\tau (\phi )}-\text{e}^{-(t_1+t_2)/\tau (\phi )}\right] .
\label{coorelation2}
\end{align}
By using slow-roll approximation~(\ref{eta}) and semi thermal equilibrium
approximation~(\ref{approx_of_equilibrium}), we can do more detailed
calculation of the second term in final equality:
\begin{eqnarray}
&&\frac{\Gamma \tau (\phi )}{(3H+\Gamma )^2(2-3H\tau (\phi ))a^3}=\frac %
\Gamma {a^3(3H+\Gamma )^2}\frac{\left( 3H+\Gamma \right) /\left(
k_p^2+V^{\prime \prime }\right) }{2-3H\left( 3H+\Gamma \right) /\left(
k_p^2+V^{\prime \prime }\right) }  \nonumber \\
&=&\frac r{(1+r)}\frac 1{a^3(3H)^2}\frac 1{{2k_p^2}/{9H^2}-3\eta -1-r}\simeq
\frac r{(1+r)}\frac{k_F}{2\bar{z}^2a^3k_F^3}=\frac{rH\bar{z}}{2(1+r)^{1/2}k^3%
},  \label{parameter_of_spectrum}
\end{eqnarray}
\end{widetext}
where we have defined a new parameter $\bar{z}=k_p/k_F=k/aH(1+r)^{\frac 12}$. Finally, the autocorrelation function reads
\begin{align}
&\left\{ \langle \delta \varphi ({\bf k},t_1)\delta \varphi ^{*}({\bf k}%
^{\prime },t_2)\rangle \right\}   \nonumber \\
&=\left[ \left\{ \delta \varphi ({\bf k},t_0)\delta \varphi ^{*}\left( {\bf %
k}^{\prime },t_0\right) \right\} \right.   \nonumber \\
&-\left. \frac{(2\pi )^3THr\bar{z}}{(1+r)^{1/2}k^3}\delta ^3\left( {\bf k}-%
{\bf k}^{\prime }\right) \right] \text{e}^{-(t_1+t_2)/\tau (\phi )}
\nonumber \\
&+\frac{(2\pi )^3THr\bar{z}}{(1+r)^{1/2}k^3}\delta ^3\left( {\bf k}-{\bf k}%
^{\prime }\right) \text{e}^{-|t_1-t_2|/\tau (\phi )}.
\label{autocorrelation_constant}
\end{align}

The correlation function is defined as
\begin{eqnarray}
    \mathcal{P}_{\delta\varphi}(\textbf{x}-\textbf{y},t_1,t_2)=\langle\delta\varphi(\textbf{x},t_1)\delta\varphi(\textbf{y},t_2)\rangle,
    \label{correlationFunction}
\end{eqnarray}
whose Fourier transformation is
\begin{eqnarray}
    \mathcal{P}_{\delta\varphi}(\textbf{k},t_1,t_2)=\int{}\frac{d^3k'}{(2\pi)^3}\langle\delta\varphi(\textbf{k},t_1)
    \delta\varphi^*(\textbf{k}',t_2)\rangle.
    \label{correlationFunctionk}
\end{eqnarray}
Applying the definition in Eq.~(\ref{correlationFunctionk}), we can simplify the autocorrelation function appeared in Eq.~(\ref{autocorrelation_constant}) as
\begin{align}
&{\cal P}_{\delta \varphi }\left( {\bf k},t_1,t_2\right)   \nonumber \\
&=\left[ {\cal P}_{\delta \varphi }\left( {\bf k},t_0,t_0\right) -\frac{k_{%
\text{B}}THr\bar{z}_{*}}{(1+r)^{1/2}k^3}\right] \text{e}^{-(t_1+t_2)/\tau
(\phi )}  \nonumber \\
&\quad+\frac{k_{\text{B}}THr\bar{z}_{*}}{(1+r)^{\frac 12}k^3}\text{e}%
^{-|t_2-t_1|/\tau (\phi )},  \label{autocorrelation_k}
\end{align}
where, without losing generality, we have set $t_1<t_2$. $\bar{z}_*$ in Eq.~\eqref{autocorrelation_k} represents the freeze-out scale that occurs at $t_1$ when $k_p(t_1)=k_F$.
The autocorrelation function $\mathcal{P}(\textbf{k},t_1,t_2)$ is obviously dependent on the initial state $\delta\varphi(\textbf{k},t_0)$. If $t_1+t_2\gg \tau(\phi)$, the memorability on initial state which appears in the first term on right hand is no longer important. Thus, the autocorrelation $\mathcal{P}_{\delta\varphi}(\textbf{k},t_1,t_2)$ is a function in terms of $|t_1-t_2|$, which means the thermal system evolves toward a stationary process.
In statistical physics, a stationary process represents a stochastic process that the variance and expectation of a system does not change when shifted in time.
If we choose an appropriate time as the initial time $t_0$ satisfied $\mathcal{P}_{\delta\varphi}(\textbf{k},t_0,t_0)={k_\text{B}THr}/{(1+r)^\frac{1}{2}k^3}$, the system of Eq.~(\ref{approx_EOM}) is totally a stationary process, by which way, we can get the power spectrum of warm inflationary scenario.
With the definition of power spectrum
\begin{eqnarray}
     P_{\delta\varphi}(\textbf{k},t)=\frac{k^3}{2\pi^2}\int{\frac{d^3k'}{(2\pi)^3}\langle\delta\varphi(\textbf{k},t_0)
     \delta\varphi(\textbf{k}',t_0)\rangle}, \label{powerspectrum}
\end{eqnarray}
it is now possible to use the definition of Eq.~(\ref{powerspectrum}) to obtain the power spectrum.
Now make average on initial state and set $t_1=t_2=t$, so correlation function can be written as
\begin{eqnarray}
{P}_{\delta \varphi }({\bf k},t) &=&\left( P_{\delta \varphi }({\bf k},t_0)-%
\frac{k_{\text{B}}THr}{2\pi ^2(1+r)^{1/2}}\right) \text{e}^{-2t/\tau (\phi )}
\nonumber \\
&&+\frac{k_{\text{B}}THr}{2\pi ^2(1+r)^{1/2}}.  \label{correlationFinal}
\end{eqnarray}

Let's first consider a condition with a strong dissipative, $r\gg 1$, which yields a scale-invariant power spectrum \cite{PhysRevD.69.083525}
\begin{eqnarray}
     P_{\delta\varphi}(\textbf{k},t)=(\Gamma H/3)^{1/2}k_\text{B}T/2\pi^2. \label{PS_strongDissi}
\end{eqnarray}
If the power spectrum of the initial state is a scale-invariant spectrum same as Eq.~\eqref{PS_strongDissi}, the spectrum $P_{\delta\varphi}(\textbf{k},t)$ is an another scale-invariant one and is totally the same as $\mathcal{P}_{\delta\varphi}(\textbf{k},t_0)$.
In other words, $P_{\delta\varphi}$ exhibits a stationary process under such a condition, which means system is already on thermal equilibrium during time interval $t>t_0$.
This method is quite similar with the analysis for correlation function of particles with Brown motion \cite{2006stme.book.....S}.
Now, consider a different condition that $P_{\delta\varphi}(\textbf{k},t_0)$ is not a scale-invariant spectrum with an arbitrary spectra index $n'$, $P_{\delta\varphi}(\textbf{k},t_0)=A(k/k_0)^{n'-1}$.
Then $P_{\delta\varphi}(\textbf{k},t)$ becomes also dependent on $k$, i.e., $P_{\delta\varphi}(\textbf{k},t)=A(k/k_0)^{n-1}$.
However, the spectra index $n$ damps slightly by tending to unit with the increasing of time.
This is an effect dominated by non-equilibrium dynamics. In this way, we can say that scalar index $n_s$ and slow-roll parameter $\beta$ are parameters that illustrate the deviation from equilibrium state during the period of cosmic inflation. Probe on cosmic microwave background shows that our Universe is almost on thermal equilibrium \cite{WMAP} if considering our Universe in early epoch as a model immersed in a thermal bath. The relation in Eq.~(\ref{correlationFinal}) indicates that $P_{\delta\varphi}(\textbf{k},t_0)$ (the initial condition of universe) becomes not so important even if we cannot give an accurate description till now.

Finally, let's have a brief conclusion on this section. From Eq.~(\ref{coorelation2}), we know that the autocorrelation function of inflaton is stationary on super-horizon scale, which means the variance or power spectrum tends to a constant during inflation at early epoch.
This result is the same as that in cold inflation. In this way, we get the power spectrum of warm inflationary scenario which shows no difference from the one via Green's function method \cite{PhysRevD.69.083525,PhysRevD.97.063516}. Similar results have been accomplished in relevant previous references \cite{PhysRevLett.74.1912,BERERA2000666}

\section{\label{generalized_Langevin_equation_WI}Warm inflation with non-Markov dissipative coefficient}

In quantum field theory, evolution equation of a field described by a differential integrate equation \cite{zee2010quantum}.
In finite temperature condition, the evolution equation of field becomes a stochastic equation of motion \cite{altland2010condensed}:
\begin{eqnarray}
     \big[\partial^2+\omega^2(x)\big]\Phi(x)+\int_0^t{\text{d}t' \Sigma(x-x')\Phi(x')}=\xi(x).\nonumber \\ \label{EOM_field}
\end{eqnarray}
From Eq.~(\ref{GFDT}), $\xi(x)$ can be interpreted as a Gaussian stochastic noise with two-point statistical correlation function
\begin{eqnarray}
     \langle\xi(x)\xi(x')\rangle=\frac{3}{2}\Sigma(|x-x'|), \label{GFDT'}
\end{eqnarray}
in which the coefficient $3/2$ comes from the three dimension of space similarly with the result in molecule statistical dynamics.
In de Sitter spacetime, we led to the following equation for the perturbed inflaton field $\delta\varphi(x)$ defined in Eq.~(\ref{inflaton}) in momentum space:
\begin{eqnarray}
     \left[\frac{\text{d}^2}{\text{d}t^2}+3H\frac{\text{d}}{\text{d}t}+\frac{\textbf{k}^2}{a^2(t)}+V''(\phi)\right]\delta\varphi(\textbf{k},t) \nonumber \\
      +\int_0^t{\text{d}t' a^3(t')\Sigma(\textbf{k};t,t')\delta\varphi(\textbf{k},t')}=\xi(\textbf{k},t). \label{GLE_inflaton0}
\end{eqnarray}
In Eq.~(\ref{GLE_inflaton0}), the integral kernel (self-energy) $\Sigma(\textbf{k};t,t')$ is a function of cosmic time $t$ and $t'$, instead of $t-t'$.
The new term, however, with conformal transformation
\begin{eqnarray}
     \bar{\Sigma}(\textbf{k};t-t')=a^{3/2}(t)a^{3/2}(t')\Sigma(\textbf{k};t,t')
\end{eqnarray}
is an integrate kernel in terms of $t-t'$. Applying the fluctuation-dissipation relation, together with the principle of general relativity, $\xi(\textbf{k},t)$ and $ \bar{\Sigma}(\textbf{k};t-t')$ follow the relation
\begin{align}
&\left\langle \xi \left( {\bf k},t\right) \xi ^{*}\left( {\bf k}^{\prime
},t^{\prime }\right) \right\rangle   \nonumber \\
=&2(2\pi )^3k_{\text{B}}T\delta ^3\left( {\bf k}-{\bf k}^{\prime }\right)
\frac{\bar{\Sigma}\left( {\bf k},|t-t^{\prime }|\right) }{%
a^{3/2}(t)a^{3/2}(t^{\prime })}.  \label{GFDT_inflaton0}
\end{align}
Define $\delta\varphi(\textbf{k},t)=a^{3/2}\tilde{\delta\varphi}(\textbf{k},t)$ and set $t=Ht$ for further treatment on Eq.(\ref{GLE_inflaton0}).
With the slow-roll approximation, then Eq.~(\ref{GLE_inflaton0}) becomes
\begin{align}
\frac{\text{d}}{\text{d}t}\tilde{\delta \varphi }({\bf k},t) &+\int_{t_0}^t{%
\text{d}t^{\prime }\gamma (t-t^{\prime })\tilde{\delta \varphi }}\left( {%
{\bf k},t^{\prime }}\right)   \nonumber \\
+ &\left[ \bar{z}^2+3\eta -\frac 32a^{-1}(t)\right] \tilde{\delta \varphi }(%
{\bf k},t)=\tilde{\xi}({\bf k},t),   \nonumber \\
\label{GLE_inflaton}
\end{align}
where $t$ is a dimensionless variable, $\gamma(t-t')\equiv \bar{\Sigma}(\textbf{k},t-t')/3H^2$, $\bar{z}\equiv k/aH=k_p/H$, and $\tilde{\xi}\equiv a^{3/2}\xi/3H^2$ with fluctuation-dissipation relation
\begin{eqnarray}
     \langle\tilde{\xi}(\textbf{k},t)\tilde{\xi}^*(\textbf{k}',t')\rangle=\frac{2(2\pi)^3k_\text{B}T}{3H^2}\delta^3(\textbf{k}-\textbf{k}')\gamma(|t-t'|).\nonumber \\ \label{GFDT_inflaton}
\end{eqnarray}
Although Eq.~(\ref{GLE_inflaton}) contains a parameter $a^{-1}$, we do not care about it too much for the reasons as follow:
\begin{itemize}
  \item The arbitrariness of initial time;
  \item When $t\gg 1/H$, this term can be neglected;
  \item In this paper, we only consider the state of system when tending to thermal equilibrium and studying the initial state will benefit us nothing.
\end{itemize}
Define $\omega^2\equiv \bar{z}^2+3\eta$ by choosing $\bar{z}$ as the value at the horizon crossing, then
\begin{equation}
\dot{\tilde{\delta \varphi }}({\bf k},t)+\int_{t_0}^t{\text{d}t^{\prime }}%
\gamma (t-t^{\prime })\tilde{\delta \varphi }({\bf k},t^{\prime })+\omega ^2%
\tilde{\delta \varphi }({\bf k},t)=\tilde{\xi}({\bf k},t).  \\
\label{GLE_inflaton_final}
\end{equation}
The process in the equation above is a stationary, non-Markov and Gaussian process.

Following the standard method used in stochastic physics \cite{klages2008anomalous}, we apply the Laplace transformation on the both sides of Eq.~(\ref{GLE_inflaton_final}), and get the solution
\footnote
{
There has the Laplace transformation relation of convolution integral
\begin{eqnarray}
    \mathcal{L}\Big[\int_0^t \text{d}\tau \beta(t-\tau)f(\tau)\Big]=\hat{\beta}(z)\hat{f}(z). \nonumber
\end{eqnarray}
}
\begin{eqnarray}
     \hat{\delta\varphi}(\textbf{k},z)=\hat{\chi}(z)\Big(\delta\varphi(\textbf{k},t_0)+\hat{\xi}(\textbf{k},z)\Big), \label{GLE_inflaton_solution0}
\end{eqnarray}
where $\delta\varphi(\textbf{k},t_0)$ is the initial value of perturbed inflaton field.
The transfer function $\hat{\chi}(z)$ is
\begin{eqnarray}
     \hat{\chi}(z) = \frac{1}{z +\hat{\gamma}(z) +\omega^2}, \label{chi}
\end{eqnarray}
with $\hat{\gamma}(z)$ being the Laplace transformation of integrate kernel
\begin{eqnarray}
     \hat{\chi}(z) = \mathcal{L}[\gamma(t)] \equiv \int_0^\infty{\text{d}t \gamma(t) \text{e}^{-zt}}. \label{LT}
\end{eqnarray}
Then applying the inverse Laplace transformation on both sides of Eq.~(\ref{GLE_inflaton_solution0}), we get the solution of Eq.~(\ref{GLE_inflaton_final}) as function of cosmic time $t$:
\begin{eqnarray}
     \tilde{\delta\varphi}(\textbf{k},t) = \chi(t)\tilde{\delta\varphi}(\textbf{k},t_0) + \int^t_0\chi(t-s)\tilde{\xi}(s)\text{d}s, \label{GLE_inflaton_solution1}
\end{eqnarray}
where $\chi(t)$ is the inverse Laplace transformation of $\hat{\chi}(z)$:
\begin{eqnarray}
     \chi(t) = \mathcal{L}^{-1}[\hat{\chi}(z)] \equiv \int_{c-\text{i}\infty}^{c+\text{i}\infty}\text{d}z\ \text{e}^{zt} \hat{\chi}(z) \label{ILT}
\end{eqnarray}
with $c>\textrm{max}\{\textrm{Re}\ \textrm{Res}[\hat{\chi}(z)]\}$ denoting the maximal one among the numbers of the real part of the residue of complex function $\hat{\chi}(z)$~\cite{davies2012integral}.

\subsection{\label{stationary_process}Proof of stationary process}

\begin{figure*}
  \centering
  \includegraphics[clip,width=0.75\textwidth]{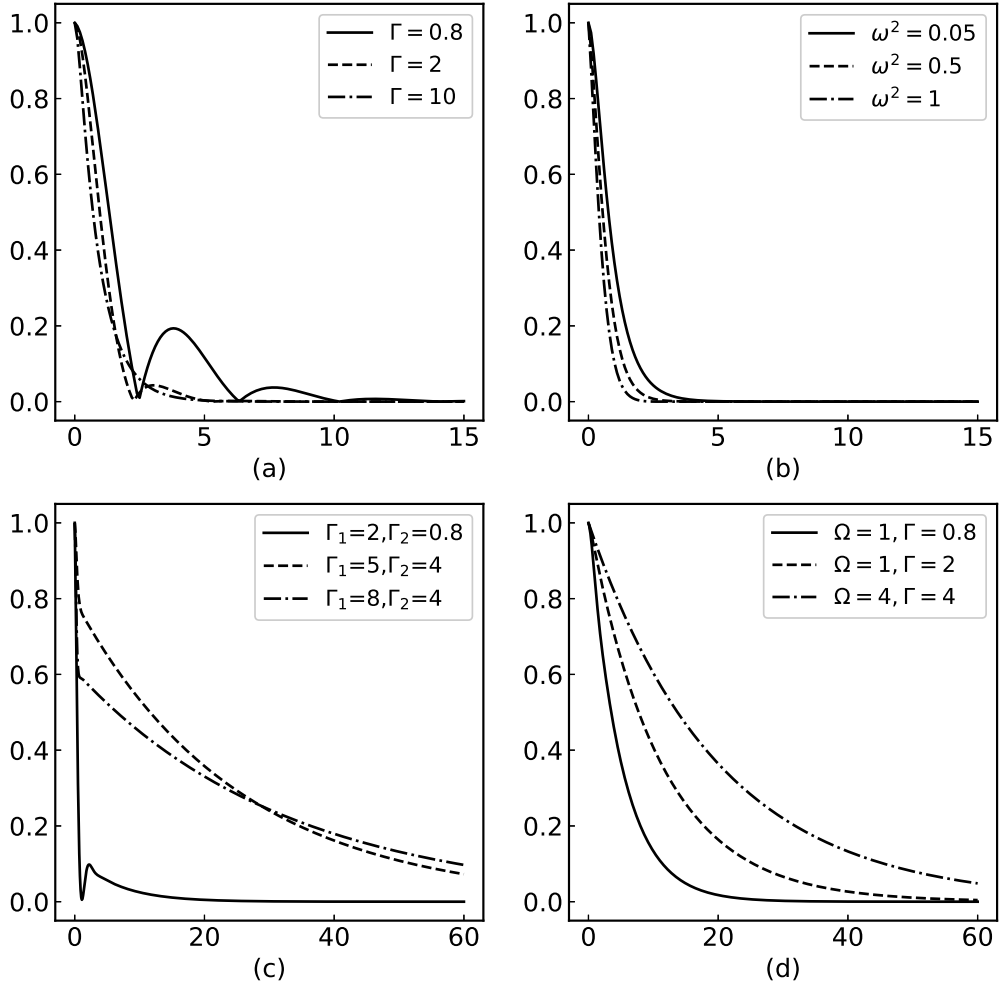}
  \caption{Different types of propagating functions $|\chi(t)|$. (a) Different modes of propagating functions $|\chi(t)|$ with red noise friction in Eq.~(\ref{damping_kernel1}), together with $\omega^2=2$ and different values of $\Gamma$ (we have set $H=1$). (b) Different modes of $|\chi(t)|$ with $\Gamma=5$ and different values of $\omega^2$. (c) Different modes of $|\chi(t)|$ with coloured noise in Eq.~(\ref{damping_kernel2}) together with the Markov friction strength $D=5$ and different combinations of $\Gamma_1$ and $\Gamma_2$. (d) Different modes of $|\chi(t)|$ with oscillating damping kernel definged in Eq.~(\ref{damping_kernel3}) and different combinations of $\Omega$ and $\Gamma$.}
  \label{damping_kernel}
\end{figure*}

As discussed in Sec.~\ref{Langevin_equation_WI}, if we want to get the power spectrum of the warm inflaton, we need to prove the system of Eq.~ (\ref{GLE_inflaton_final}) satisfying a stationary process. The autocorrelation function of the warm inflaton is
\begin{widetext}
\begin{align}
    & \ \ \{ \langle \tilde{\delta\varphi}(\textbf{k}, t_1)\tilde{\delta\varphi}^*(\textbf{k}', t_2) \rangle \}  \nonumber \\
    & = \chi(t_1)\big\{\tilde{\delta\psi}(\textbf{k}, t_0)\tilde{\delta\varphi}(\textbf{k}', t_0) \big\}{\chi}^*(t_2) + \int^{t_1}_0\int^{t_2}_0\chi(t_1-s_1)\chi^*(t_2-s_2)\langle\tilde{\xi}(s_1)\tilde{\xi}^*(s_2)\rangle \text{d}s_1 \text{d}s_2 \nonumber \\
    & = \chi(t_1)\big\{\tilde{\delta\psi}(\textbf{k}, t_0)\tilde{\delta\varphi}(\textbf{k}', t_0) \big\}{\chi}^*(t_2) + \frac{2(2\pi)^3k_\text{B}T\delta^3(\textbf{k}-\textbf{k}')}{3H^2}\int^{t_1}_0\int^{t_2}_0\chi(t_1-s_1)\chi^*(t_2-s_2)\gamma(|s_1-s_2|) \text{d}s_1 \text{d}s_2.   \nonumber \\
    \label{autocorrelation_GLE0}
\end{align}
The double integration in Eq.~(\ref{autocorrelation_GLE0}) could be calculated by performing the double Laplace transforation \cite{FOX1978179,1977JMP....18.2331F,1977JSP....16..259F}
\begin{eqnarray}
     & &\quad \int^\infty_0 \text{d}t_1\int^\infty_0\text{d}t_2 \text{e}^{-z_1t_1}\text{e}^{-z_2t_2}\int^{t_1}_0 \text{d}s_1\int^{t_2}_0\text{d}s_2\chi(t_1-s_1)\chi^*(t_2-s_2)\gamma(|s_1-s_2|) \nonumber \\
     & & = \int^\infty_0 \text{d}s_1\int^\infty_{s_1}\text{d}t_1\int^\infty_0 \text{d}s_2\int^\infty_{s_2}\text{d}t_2 \text{e}^{-z_1(t_1-s_1)}\text{e}^{-z_2(t_2-s_2)}\chi(t_1-s_1)\chi^*(t_2-s_2)\text{e}^{-z_1s_1}\text{e}^{-z_2s_2}\gamma(|s_1-s_2|) \nonumber \\
     & & = \int^\infty_0 \text{d}s_1\int^\infty_0\text{d}\tau_1\int^\infty_0 \text{d}s_2\int^\infty_0\text{d}\tau_2 \text{e}^{-z_1\tau_1}\text{e}^{-z_2\tau_2}\chi(\tau_1)\chi^*(\tau_2)\text{e}^{-z_1s_1}\text{e}^{-z_2s_2}\gamma(|s_2-s_1|) \nonumber \\
     & & = \hat{\chi}(z_1)\hat{\chi}^*(z_2)\int^\infty_0\text{d}s_2\int^\infty_0\text{d}s_1\text{e}^{-z_1s_1}\text{e}^{-z_2s_2}\gamma(|s_2-s_1|). \label{autocorrelation_GLE1}
\end{eqnarray}
The last double integration in Eq.~(\ref{autocorrelation_GLE1}) also contains a double Laplace transform,
\begin{eqnarray}
    & &\quad \int^\infty_0\text{d}s_2\int^\infty_0\text{d}s_1e^{-z_1s_1}e^{-z_2s_2}\gamma(|s_2-s_1|) \nonumber \\
    & & = \left(\int^\infty_0\text{d}s_2\int^\infty_{s_2}\text{d}s_1 + \int^\infty_0\text{d}s_1\int^\infty_{s_1}\text{d}s_2 \right) \text{e}^{-z_1s_1}\text{e}^{-z_2s_2}\gamma(|s_2-s_1|) \nonumber \\
    & & = \int^\infty_0\text{d}s_2\int^\infty_0\text{d}\tau \text{e}^{-(z_1+z_2)s_2}\text{e}^{-z_1\tau}\gamma(\tau)
    +\int^\infty_0\text{d}s_1\int^\infty_0\text{d}\tau'\text{e}^{-(z_1+z_2)s_1}\text{e}^{-z_2\tau'}\gamma(\tau')\nonumber \\
    & & =\frac{\hat{\gamma}(z_1)+\hat{\gamma}(z_2)}{z_1+z_2}. \label{autocorrelation_GLE2}
\end{eqnarray}
In second equality, we separate the integration into two parts: the integration on region $s_1>s_2$ and the integration on region $s_2>s_1$. Substituting Eq.~(\ref{autocorrelation_GLE2}) into Eq.~(\ref{autocorrelation_GLE1}), we have
\begin{equation}
\quad \int_0^\infty \text{d}t_1\int_0^\infty \text{d}t_2\text{e}^{-z_1t_1}%
\text{e}^{-z_2t_2}\int_0^{t_1}\text{d}s_1\int_0^{t_2}\text{d}s_2\chi
(t_1-s_1)\chi ^{*}(t_2-s_2)\gamma (|s_1-s_2|)=\hat{\chi}(z_1)\hat{\chi}%
^{*}(z_2)\frac{\hat{\gamma}(z_1)+\hat{\gamma}(z_2)}{z_1+z_2}
\label{autocorrelation_GLE3}
\end{equation}
According to Eq.~(\ref{chi}), it follows the relation
\begin{eqnarray}
    \hat{\chi}(z_1) \hat{\gamma}(z_1) = \frac{\hat{\gamma}(z_1)}{z_1+\hat{\gamma}(z_1)+\omega^2} = 1 - \hat{\chi}(z_1)(z_1 +\omega^2). \label{autocorrelation_GLE4}
\end{eqnarray}
We should notice that $\gamma(t-s)$ is a symmetry function (matrix), while $\omega^2=-\text{i}\Omega$ is an asymmetry parameter (matrix) \cite{FOX1978179}, which leads to
\begin{eqnarray}
     \hat{\chi}^*(z_2)\hat{\gamma}(z_2) = 1- \hat{\chi}^*(z_2)(z_2-\omega^2). \label{autocorrelation_GLE5}
\end{eqnarray}
Then we have
\begin{eqnarray}
    \hat{\chi}(z_1)\hat{\chi}^*(z_2)\frac{\hat{\gamma}(z_1)+\hat{\gamma}(z_2)}{z_1+z_2} = \frac{\hat{\chi}(z_1)+\hat{\chi}^*(z_2)}{z_1+z_2}-\hat{\chi}(z_1)\hat{\chi}^*(z_2). \label{autocorrelation_GLE6}
\end{eqnarray}
Based on the converse calculation of Eq.~(\ref{autocorrelation_GLE2}), the first term in Eq.(\ref{autocorrelation_GLE6}) follows the inversion Laplace transform relation
\begin{eqnarray}
    \mathcal{L}^{-1}\Big[ \frac{\hat{\chi}(z_1) +\hat{\chi}^*(z_2)}{z_1+z_2} \Big]\equiv \tilde{\chi}(|t_1-t_2|) = \theta(t_1-t_2)\chi(t_1-t_2) +\theta(t_2-t_1)\chi^*(t_2-t_1). \label{autocorrelation_GLE7}
\end{eqnarray}
Applying Eqs.~(\ref{autocorrelation_GLE0}),~(\ref{autocorrelation_GLE3})~and~(\ref{autocorrelation_GLE7}), the autocorrelation function of $\tilde{\delta\varphi}(\textbf{k}, t)$ is given by
\begin{align}
    &\ \ \{ \langle \tilde{\delta\varphi}(\textbf{k}, t_1)\tilde{\delta\varphi}^*(\textbf{k}', t_2) \rangle \}  \nonumber \\
    & = \left[\{\tilde{\delta\varphi}(\textbf{k}, t_0)\tilde{\delta\varphi}(\textbf{k}', t_0)\}
    -\frac{2(2\pi)^3k_\text{B}T\delta^3(\textbf{k}-\textbf{k}')}{3H^2}\right]\chi(t_1)\chi^*(t_2) \nonumber \\
    & \quad+\frac{2(2\pi)^3k_\text{B}T\delta^3(\textbf{k}-\textbf{k}')}{3H^2}\Big[\theta(t_1-t_2)\chi(t_1-t_2) +\theta(t_2-t_1)\chi^*(t_2-t_1) \Big].
    \label{autocorrelation_GLE_final}
\end{align}
\end{widetext}
If choosing an appropriate initial time such that the variance on $\tilde{\delta\varphi}(\textbf{k}, t_0)$ satisfies the relation
\begin{eqnarray}
    \langle \tilde{\delta\varphi}(\textbf{k}, t_0)\tilde{\delta\varphi}^*(\textbf{k}', t_0) \rangle = \frac{2(2\pi)^3k_\text{B}T\delta^3(\textbf{k}-\textbf{k}')}{3H^2},
\end{eqnarray}
the autocorrelation function is a function in terms of variable $|t_1-t_2|$, which exhibits the stationarity of the process.
It's worth noting that the first term on the right-hand of Eq.~(\ref{autocorrelation_GLE_final}) contains two functions dependent on $t_1$ and $t_2$, instead of $|t_1-t_2|$. but the term including $\chi(t_1)\chi^*(t_2)$ contributes a dramatically damping trend to autocorrelation function. As a brief illustration, the first panel in Fig.~\ref{damping_kernel} plots the portraits of $|\chi(t)|$ with different values of $\Gamma$ (set $H=1$) in a simple memory kernel
\begin{eqnarray}
    \gamma(t-t') = \Gamma\text{e}^{-\Gamma |t-t'|}.  \label{damping_kernel1}
\end{eqnarray}
Together, the second panel includes three lines with different values of $\omega^2$. In third panel, the damping kernel function is
\begin{eqnarray}
    \gamma(t-t') = D\big[\Gamma_1\text{e}^{-\Gamma_1 |t-t'|}-\Gamma_2\text{e}^{-\Gamma_2 |t-t'|}\big],  \label{damping_kernel2}
\end{eqnarray}
where $D$ is called Markov friction strength. The damping kernel function above describes a noise whose spectrum density function vanishes at both low and high frequency. Another damping kernel is an oscillating damping mode \cite{PhysRevD.62.083517}
\begin{eqnarray}
    \gamma(t-t') = \frac{\Omega\cos(2\Omega |t-t'|)+\Gamma\sin(2\Omega |t-t'|)}{2\Omega^2(\Gamma^2+\Omega^2)}\text{e}^{-2\Gamma |t-t'|}. \nonumber \\ \label{damping_kernel3}
\end{eqnarray}

\subsection{\label{Power_spectrum}Power spectrum}
With the proof on stationary process, now we compute the power spectrum of Eq.~(\ref{GLE_inflaton_final}).
Performing the derivative of Eq.~(\ref{ILT}) with respect to $t$, there has
\begin{align}
    \dot{\chi}(t) & = \int_{c-\text{i}\infty}^{c+\text{i}\infty}\text{d}z\ \frac{z}{z +\hat{\Gamma}(z) +\omega^2}\text{e}^{zt}\nonumber \\
    & = -\omega^2\chi(t) -\int^t_0\chi(t-s)\gamma(s)\text{d}s.  \label{derivative_chi}
\end{align}

Define a new stochastic perturbation variable
\begin{align}
    y(t) & = \tilde{\delta\phi}(\textbf{k},t) - \chi(t)\tilde{\delta\phi}(\textbf{k},t_0) \nonumber \\
     & = \int^t_0\chi(t-s)\tilde{\xi}(s)\text{d}s. \label{y}
\end{align}
The variance of $y$ is
\begin{align}
    A(t) \equiv \langle y(t)\bar{y}(t)\rangle = &  \int^t_0\int^t_0\chi(t-s_1)\chi^*(t-s_2) \nonumber \\
     & \times \langle\tilde{\xi}(s_1)\tilde{\xi}^*(s_2)\rangle \text{d}s_1 \text{d}s_2. \label{A}
\end{align}
By using Eq.~(\ref{GFDT_inflaton}), we have
\begin{align}
    A(t) = & \frac{2(2\pi)^3k_\text{B}T\delta^3(\textbf{k}-\textbf{k}')}{3H^2}\nonumber \\
    & \times \ \int^t_0\int^t_0\chi(t-s_1)\chi^*(t-s_2)\gamma(|s_1-s_2|)\text{d}s_1 \text{d}s_2 \nonumber \\
    = & \ \frac{2(2\pi)^3k_\text{B}T\delta^3(\textbf{k}-\textbf{k}')}{3H^2} \nonumber \\
    & \times \int^t_0\int^t_0\chi(\tau)\gamma(|\tau-\tau'|){\chi}^*(\tau')\text{d}\tau \text{d}\tau'. \label{variance_A1}
\end{align}
Perform the derivative with respect to $t$ of $A(t)$,
\begin{align}
    \dot{A}(t) & =\frac{4(2\pi)^3k_BT\delta^3(\textbf{k}-\textbf{k}')}{3H^2}{\chi}^*(t)\int^t_0\gamma(t-\tau')\chi(\tau)\text{d}\tau \nonumber \\
    & =\frac{4(2\pi)^3k_\text{B}T\delta^3(\textbf{k}-\textbf{k}')}{3H^2}{\chi}^*(t)\mathcal{L}^{-1}[\gamma(z)\chi(z)] \nonumber \\
    & =-\frac{4(2\pi)^3k_\text{B}T\delta^3(\textbf{k}-\textbf{k}')}{3H^2}\big[\chi^*(t)\dot{\chi}(t)+\omega^2|\chi(t)|^2\big] \nonumber \\
    & =-\frac{2(2\pi)^3k_\text{B}T\delta^3(\textbf{k}-\textbf{k}')}{3H^2}\left[\frac{\text{d}}{\text{d}t}|\chi(t)|^2+2\omega^2|\chi(t)|^2\right].\nonumber \\
\end{align}
Thus, the variance of $y$ reads
\begin{eqnarray}
    A(t) = \frac{2(2\pi)^3k_BT\delta^3(\textbf{k}-\textbf{k}')}{3H^4}\left[1-|\chi(t)|^2-2\omega^2 B(t)\right], \nonumber\\ \label{A_final}
\end{eqnarray}
since $A(0)=0,\chi(0)=1$. The expression of $B(t)$ in Eq.~\eqref{A_final} is
\begin{eqnarray}
    B(t) = \int^t_0 |\chi(t)|^2 \text{d}t' \label{B}
\end{eqnarray}
with $B(0)=0$. Based on the calculations above, the autocorrelation function writes
\begin{align}
    & \ \ \{ \langle \tilde{\delta\varphi}(\textbf{k}, t)\tilde{\delta\varphi}^*(\textbf{k}', t) \rangle \}  \nonumber \\
    & = \Big(\{\tilde{\delta\varphi}(\textbf{k}, t_0)\tilde{\delta\varphi}(\textbf{k}', t_0)\}
    -\frac{2(2\pi)^3k_\text{B}T\delta^3(\textbf{k}-\textbf{k}')}{3H^2}\Big)|\chi(t)|^2 \nonumber \\
    & \quad+\frac{2(2\pi)^3k_\text{B}T\delta^3(\textbf{k}-\textbf{k}')}{3H^2}\Big[1-2\omega^2 B(t)\Big].
    \label{spectrum}
\end{align}
We see that the first term on the right-hand of Eq.~(\ref{spectrum}) contains two functions dependent on cosmic time $t$. Although this term is time dependent, it could almost be ignored under lager scale limit $\bar{z}=k/aH\ll 1$ and slow-roll condition $\eta\ll 1+r$ ($\omega^2\ll 1+r$). Besides, the first panel and the second panel in Fig.~\ref{damping_kernel} show that the portrait of $|\chi(t)|^2$ has a sharp distribution near $t=0$ with $\Gamma/H\gg 1$.
The integral on $|\chi(t)|^2$ does not arise a large value and is even much smaller than unit, i.e., $B(+\infty)\ll 1$. So $1-2\omega^2 B(t)$ can be regarded as unit since both $\omega^2$ and $B(t)$ are much smaller than the unit. From the discussions in this section, the memory kernel $\gamma(t)$ drives the system evolving to the equilibrium state, while $\omega^2$ compels the system deviating from the equilibrium state, which means the slow-roll parameter $\eta$ must be much smaller than the unit. According to the definition of Eq.~(\ref{powerspectrum}), we finally get the power spectrum of $\delta\varphi(\textbf{k}, t)$ at horizon crossing:
\begin{eqnarray}
    P_{\delta\varphi}(\textbf{k},t)\simeq\frac{k_\text{B}TH}{3\pi^2H^3a^3k^{-3}}=\frac{k_\text{B}TH}{3\pi^2},    \label{spectrum_of_GLE}
\end{eqnarray}
which is quite analogous to the one in cold inflation.
This result is based on the fluctuation-dissipation relation of Eqs.~(\ref{GFDT})~and~(\ref{GFDT_inflaton}). Following the relation of Eq.~(\ref{GFDT'}) and repeating the computations in this section, we obtain another power spectrum of warm inflaton
\begin{eqnarray}
    P_{\delta\varphi}(\textbf{k},t)\simeq \frac{H^2}{4\pi^2},    \label{spectrum_of_GLE'}
\end{eqnarray}
which is exactly the same as the one in cold inflation \cite{2003moco.book.....D}!

\subsection{\label{approximate_condition}Approximate condition}

As usual, cold inflationary model needs two approximate parameters $\varepsilon$ and $\eta$ as seen in Eqs.~(\ref{varepsilon})~and~(\ref{eta}).
The warm inflationary model with Langevin scenario also need another approximate parameter $\beta$ as discussed in Sec.~\ref{Langevin_equation_WI}, where the thermal equilibrium approximation requires that the relaxation time $\tau(\psi)$ is much smaller than the inverse of the expansion rate $3H$.
This parameter describes the departure of the thermodynamic system from its equilibrium state.
The generalized Langevin scenario, however, does not includes a parameter like relaxation time appeared in Eq.~(\ref{approx_solution}) because there is a integral in the stochastic differential equation~(\ref{GLE_inflaton_final}). Fortunately, we still have another way to give the approximation in generalized Langevin warm inflation. The statistical physics theory has shown that a Langevin equation generates a Fokker-Planck equation that is partial differential equation to describe the time evolution of the probability density function (also called two time distribution function) of a particle under the influence of attracting potential and random forces \cite{p2012langevin,OLIVARESRIVAS201676,1976JChPh..64..124A}.
Since the stochastic differential equation~(\ref{GLE_inflaton_final}) exhibits a Gaussian process as introduced in the previous of this section, the probability density function is given by
\begin{eqnarray}
    W(y,y_0;t) = \Big| \frac{1}{2\pi A(t)}\Big|^{\frac{1}{2}}\exp\left\{-\frac{1}{2}y(t)A^{-1}(t)y^*(t)\right\},     \label{pdf}
\end{eqnarray}
where $y(t)$ and $A(t)$ are defined in Eqs. (\ref{y}) and (\ref{A}) respectively.
The PDF $W(y,y_0;t)$ follows the Fokker-Planck equation as
\begin{align}
    \frac{\partial}{\partial t}W(y,y_0;t) = & -\frac{\dot{\chi(t)}}{\chi(t)} \frac{\partial}{\partial y}\big(yW(y,y_0;t) \big) \nonumber \\
    & -\frac{k_\text{B}TH}{3}\frac{\dot{\chi(t)}}{\chi(t)}\frac{\partial^2}{\partial y^2}W(y,y_0;t).     \label{FP_equation}
\end{align}
In the equation above, the parameter $-{\dot{\chi(t)}}/{\chi(t)}$ is just the inversion of relaxation time $\tau_\text{rex}$.
Similarly with the discussion in Sec.~\ref{generalized_Langevin_equation_WI}, the relaxation time must be much smaller than the Hubble time, i.e., $\tau_\text{rex}\ll 1/3H$.

We still, however, have not given the analytic expression of the approximation condition till now.
Using the slow-roll condition, we can directly consider the propagating function $\chi(t)$ mainly contributed from the only term of integral kernel.
Concretely, we set the damping kernel proportional to a exponential function
\begin{eqnarray}
    \gamma(t) = \bar{\gamma}(t)\Gamma\text{e}^{-\Gamma t},
\end{eqnarray}
where $\bar{\gamma}(t)$ is a slow variation function namely $\dot{\bar{\gamma}}/\bar{\gamma}\Gamma\ll 1$.
Thus, the propagating function reads
\begin{eqnarray}
    \chi(t)\approx \mathcal{L}^{-1}\big\{\big[\Gamma(z)\big]^{-1}\big\} \approx \bar{\gamma}(t)\Gamma^2\text{e}^{-\Gamma t}.   \label{approx_chi}
\end{eqnarray}
If $\Gamma\gg H$, there approximately exists
\begin{eqnarray}
    \gamma(t) \approx -\bar{\gamma}(t)\delta'(t),
\end{eqnarray}
which leads to the Langevin equation~(\ref{EOM_pertubed_field}).
Using Eq.~(\ref{approx_chi}) and $\tau_\text{rex}\ll 1/3H$, we have
\begin{eqnarray}
    \frac{\dot{\bar{\gamma}}}{3H\bar{\gamma}}\ll 1+r. 
\end{eqnarray}
Specially, if $\bar{\gamma}(t)$ is a function as the average background inflaton field $\phi(t)$, i.e., $\bar{\gamma}(t)=\Upsilon[\phi(t)]$, by applying the slow-roll condition, we also obtain
\begin{eqnarray}
    \beta &=& \frac{1}{8 \pi G}\frac{\Upsilon,_\phi V,_{\phi}}{\Gamma V} \ll 1+r , \label{beta'}
\end{eqnarray}
which is just the slow-roll approximation~(\ref{beta}).

\section{\label{conclusion}Conclusion and discussion}

In this paper, we discuss the Markovian and non-Markovian statistical dynamical problem of the warm inflationary scenario via a Langevin language.
In Sec. \ref{Langevin_equation_WI}, we study a simple condition with constant dissipative coefficient. In this model, if we reckon the initial state which have been already on a thermal equilibrium state, the perturbed inflaton field exhibits a stationary process on superhorizon sclae. In other words, the variance of the perturbed field does not change with the time on large scale, which is similar to the cold inflationary scenario.
If the initial state is not on equilibrium, the variance tends to a constant as an exponentially damping form, and this constant is the spectrum of perturbed warm inflaton field. The non-equilibrium initial state leads to a time dependent spectrum which corresponds to the spectrum index. Using the semi-thermal equilibrium approximation, we also derive a freeze-out scale in warm inflation which is always smaller than that in cold inflation.

In Sec. \ref{generalized_Langevin_equation_WI}, we study the warm inflation with a generalized Langevin equation scenario as a stochastic differential integral equation in which the dissipative effect is described by a time dependent integral kernel as long as the initial state is on equilibrium. We also prove that the stochastic process is also a stationary process. With the general fluctuation-dissipation theory, we derive the power spectrum of the perturbed field as well, but it is a time dependent one. If we consider the large scale limit and the slow-roll approximation, we can reckon that this power spectrum is also time independent. So, in this method, we get a scale-invariant power spectrum, which is quite analogous to the one in cold inflation. If we choose the fluctuation-dissipation relation of Eq.(\ref{GFDT'}), we obtain a scale-invariant power spectrum which is the same as that in cold inflation. These results show us that the warm inflation model is a extremely possible scenario to substitute the cold inflation model. As the discussion on warm inflation with Langevin equation, we also treat the early universe satisfies the semi-thermal equilibrium condition. As a result, this condition leads to a approximation on damping kernel analogous to Eq.(\ref{beta}).

With the discussion above, we strongly believe that the warm inflationary model is a alternative scenario to cold inflation. In this paper, we only show a brief picture to illustrate the rationality of explanation of the warm inflation via statistical physics method. There are still many questions which deserve further discussion, such as the initial condition problem, the spectrums from different choice on potential function $V(\phi)$ and so on.

\begin{acknowledgments}
This work was supported by the National Natural Science Foundation of China (Grants No. 11575270, No. 11175019, No. 11235003, No. 11675018, No. 11790325, and No. 11735005).
\end{acknowledgments}

\end{document}